\documentclass[fleqn,usenatbib]{mnras}

\usepackage[T1]{fontenc}
\usepackage{ae,aecompl}

\usepackage{color}
\usepackage{natbib}
\usepackage[dvipsnames]{xcolor}
\usepackage{flushend}
\usepackage{bm}

\usepackage{graphicx}	
\usepackage{amsmath}	
\usepackage{amssymb}	

\DeclareMathAlphabet{\mathcal}{OMS}{cmsy}{m}{n}
\DeclareMathAlphabet\mathbfcal{OMS}{cmsy}{b}{n}

\newcommand{\de}{{\rm d}}
\newcommand{\lcdm}{$\Lambda$CDM}
\newcommand{\om}{\Omega_{\rm m}}
\newcommand{\ob}{\Omega_{\rm b}}

\newcommand{\odm}{\Omega_{\rm DM}}
\newcommand{\As}{\mathcal A_{\rm s}}
\newcommand{\ns}{{n_{\rm s}}}
\newcommand{\fnl}{f_{\rm NL}}
\newcommand{\ho}{H_0}

\newcommand{\lm}{\ell_{\rm min}}
\newcommand{\lM}{\ell_{\rm max}}
\newcommand{\planck}{\textit{Planck}}

\def\be{\begin{equation}}
\def\ee{\end{equation}}

\title[Optimal spectroscopic angular power spectra]{Optimised angular power spectra for spectroscopic galaxy surveys}

\author[S. Camera {\rm et al.}]{Stefano Camera,$^{1,2,3,4,5}$\thanks{E-mail: stefano.camera@unito.it.}
Jos\'e Fonseca,$^{4}$
Roy Maartens$^{4,6}$
\& M\'ario G. Santos$^{4,7}$
\\
$^{1}$Dipartimento di Fisica, Universit\`a degli Studi di Torino, via P. Giuria 1, 10125 Torino, Italy\\
$^{2}$INFN -- Istituto Nazionale di Fisica Nucleare, Sezione di Torino, via P. Giuria 1, 10125 Torino, Italy\\
$^{3}$INAF -- Istituto Nazionale di Astrofisica, Osservatorio Astrofisico di Torino, strada Osservatorio 20, 10025 Pino Torinese, Italy\\
$^{4}$Department of Physics \& Astronomy, University of the Western Cape, Cape Town 7535, South Africa\\
$^{5}$Jodrell Bank Centre for Astrophysics, The University of Manchester, Alan Turing Building, Oxford Road, Manchester M13 9PL, UK\\
$^{6}$Institute of Cosmology \& Gravitation, University of Portsmouth, Dennis Sciama Building, Portsmouth PO1 3FX, UK\\
$^{7}$SKA South Africa, The Park, Cape Town 7405, South Africa}

\date{Accepted XXX. Received YYY; in original form ZZZ}

\pubyear{2018}

\begin{document}
\label{firstpage}
\pagerange{\pageref{firstpage}--\pageref{lastpage}}
\maketitle

\begin{abstract}
The angular power spectrum $C_\ell(z_i,z_j)$ is a gauge-independent observable that is in principle the natural tool for analysing galaxy number counts. In practice, the problem is that the computational requirements for next-generation spectroscopic surveys such as \textit{Euclid} and the Square Kilometre Array are currently unfeasible. We propose a new method to save computational time for spectroscopic angular power spectra. This hybrid method is modelled on the Fourier power spectrum approach of treating relatively thick redshift bins ($\Delta z \sim 0.1$) as separate surveys. In the hybrid method, each thick bin is further subdivided into thin bins ($\delta z \sim 0.01$); all the correlations within each thick bin are computed, while cross-bin correlations beyond the thick bins are neglected. Constraints on cosmological parameters from the hybrid method are comparable to those from the standard galaxy $P_{\rm g}({\bm k},z)$ analysis---but they have the advantage that cosmic evolution, wide-angle and lensing effects are naturally included, while no Alcock-Paczynski correction is needed. The hybrid method delivers much tighter constraints than a 2D tomographic approach that is typical for photometric surveys, which considers only thick bins and the correlations between them. Furthermore, for standard cosmological parameters our method is not biased by neglecting the effects of lensing on number counts, while the tomographic method is strongly biased.
\end{abstract}


\section{Introduction}\label{sec:introduction}
The concordance cosmological model \lcdm\ represents the current best fit to a number of very different observations across a wide range of physical scales and cosmic time. Temperature anisotropies of the cosmic microwave background measured by the \planck\ satellite \citep{Ade:2015xua}, combined with many other data sets, have shown that present-day data does not favour any extension to \lcdm. Nevertheless, key questions remain to be answered, like what drives the late-time cosmic accelerated expansion, or whether there is non-Gaussianity in the primordial density fluctuations. Furthermore, some tensions among data sets---in particular, between high- and low-redshift observables---have been claimed to herald breaches in the adamantine \lcdm\ model \citep{Spergel:2013rxa,Addison:2015wyg,Raveri:2015maa, Joudaki:2016mvz,Battye:2014qga,Joudaki:2016kym,Charnock:2017vcd,Pourtsidou:2016ico,An:2017crg,Camera:2017tws}. All of this calls for a better understanding of the late-time Universe and the growth and clustering of cosmic structures, from ultra-large to mildly nonlinear scales.

The main envisaged methods to probe the large-scale structure are measurements of cosmic shear and the creation of huge galaxy catalogues with which we can reconstruct the clustering of dark matter haloes over a wide range of scales and redshifts. In this paper, we focus on the latter method, namely on the measurement of the power spectrum of galaxy number counts, although we will comment on cosmic shear in the conclusions (\S~\ref{sec:conclusions}). Up to now, power spectrum measurements have mainly relied on two approaches: reconstructing baryon acoustic oscillation (BAO) peaks and redshift-space distortions (RSD) from the three-dimensional (3D) galaxy positions in a spectroscopic redshift survey, or estimating the 2D clustering of galaxies {in redshift slices} from broad-band photometric measurements. Despite the fact that photometric galaxy catalogues are significantly more populated than spectroscopic ones, the former method is more informative, since it gives access to the cosmological information encoded in the fully 3D cosmic web. However, it relies on a number of assumptions that require various add-on techniques in order to exploit the data quality and sky and redshift coverage envisaged by next-generation experiments, as we discuss in \S~\ref{ssec:3D}.

In this paper, we introduce a new method to optimise angular power spectrum computations for spectroscopic galaxy surveys. This method is in some sense a combination of the galaxy Fourier power spectrum $P_{\rm g}({\bm k},z)$ approach (discussed in \S~\ref{ssec:3D}) and a 2D tomographic approach (outlined in \S~\ref{ssec:2D}). For this reason, we refer to the new method as `hybrid'. In \S~\ref{sec:results}, we present our main results, showing how our hybrid approach yields constraints on cosmological parameters that are comparable to those from a standard $P_{\rm g}({\bm k},z)$ analysis, and are more than twice as tight as those obtained with a standard tomographic analysis. This is further supported by a study of the `information gain' earned by going from the standard tomography to the hybrid method. 

We also demonstrate that our hybrid approach is robust with respect to not including the corrections to galaxy number counts due to gravitational lensing. Specifically, we show that best-fit values of cosmological parameters estimated via density fluctuations and RSD alone---namely, neglecting lensing---are biased by less than 20\% of the 1$\sigma$ error on the parameter.\footnote{Note that this does not apply to the non-Gaussianity parameter $\fnl$, whose best-fit is biased by neglecting lensing (see \S~\ref{ssec:PNG}).} This is important, since the numerical computation of the lensing correction, being an integrated effect, is significantly slower. Hence, our method allows for faster computation of the data likelihood and is then more suitable for implementation in Monte Carlo Markov Chain pipelines, relevant in view of upcoming spectroscopic galaxy surveys, such as the European Space Agency \textit{Euclid} satellite \citep{Laureijs:2011gra,Amendola:2012ys,Amendola:2016saw} or the Square Kilometre Array (SKA) \citep{Maartens:2015mra,Abdalla:2015zra}. Since our method follows one key aspect of the $P_{\rm g}({\bm k},z)$ approach, i.e.\ treating thick bins as separate surveys, this suggests that the Fourier power spectrum method might not be biased by excluding lensing effects.

\section{Methodology}
\subsection{Statistical tools}
We consider a 6-dimensional cosmological parameter set $\boldsymbol\vartheta=\big\{\ob,\odm,\ns,\As,H_0\big\}$, with $\ob$ and $\odm$ the baryonic and dark matter density fractions, $\ns$ and $A_{\rm s}$ respectively the slope and the amplitude of the primordial power spectrum measured at some pivot scale $k_0$, with $\As=\ln(10^{10}A_{\rm s})$, and $\ho=100h\,\mathrm{km/s/Mpc}$ the Hubble constant. Fiducial values for the parameters are $\overline{\boldsymbol\vartheta}=\{0.05,0.26,0.9667,3.06,67.74\}$.

We work in a Bayesian statistics framework, and make use of Fisher matrices to forecast the capabilities of future surveys \citep[e.g.][]{Trotta:2008qt}. All relevant formulas can be found in Appendix~\ref{app:Fisher}, together with a description of the tests we performed to ensure the stability of the matrices and the robustness of our results. We use `relative errors' to denote marginal errors $\sigma(\vartheta_\alpha)$ divided by parameter fiducial values $\overline{\vartheta}_\alpha$, and `relative biases' on parameters to mean biases on cosmological parameters in units of the measurement precision, i.e.\ $b(\vartheta_\alpha)/\sigma(\vartheta_\alpha)$.

\subsection{Observational assumptions}\label{obas}
To compare results from our optimised hybrid method to the standard 2D tomographic approach, we apply both methods to the same observational set-up, namely a Stage IV Dark Energy Task Force (DETF) cosmological experiment \citep[see][]{Albrecht:2006um}. To compute the redshift distribution of sources and magnification bias, we use model 3 of \citet[]{Pozzetti:2016cch}. For this work we assume a flux cut threshold for the survey $\mathcal F_*=3\times 10^{-16}$ erg s$^{-1}$ cm$^{-2}$ in the redshift range $0.6\le z\le2$. Using this assumption we find the following fits for the spectroscopic distribution of sources per solid angle per unit redshift
\begin{multline}
\de N_{\rm g}/\de z=z^{1.281}\\\times\exp\left(9.976-2.317\,z-0.617\,z^2+0.265\,z^3-0.030\,z^4\right),\label{eq:ng}
\end{multline}
and for the magnification bias
\begin{equation}
\mathcal Q=0.66+2.95\,z-1.59\,z^2+0.40\,z^3-0.04\,z^4,
\end{equation}
where sometimes the magnification bias is referred to as $s$, and $\mathcal Q=5s/2$ holds. Note that these fits are only valid for $z\in[0.6,3]$ and do not reproduce well the results out of this range. We model the galaxy bias as $b_{\rm g}(z)=\sqrt{1+z}$ \citep{Amendola:2012ys,Amendola:2016saw}, and assume a sky area of $ 15,000\,{\rm deg}^2$. All angular power spectra are computed with the public code \texttt{CAMB\_sources} \citep{Challinor:2011bk} using the distribution of sources given by Eq.~\eqref{eq:ng}.

When we quote results for standard tomography (see \S~\ref{ssec:2D}), we divide the source redshift distribution into 20 equi-populated, top-hat redshift bins, with edges smeared by a Gaussian window with $\sigma_{\rm z}=0.002$. For our hybrid method, we follow the recipe outlined in \S~\ref{ssec:2.5D}.

As a final remark, we employ the linear matter power spectrum. Depending on the maximum angular wavenumber considered, this may not be correct at low redshifts, since nonlinear scales could already be contributing to the angular power spectrum. This can be seen via the Limber approximation, valid for small angles, $k=\ell/\chi$. At low redshift,  $\lM=800$ already exceeds the nonlinear threshold $k_{\rm nl}\simeq0.2\,h/\mathrm{Mpc}$. Nonetheless, our aim is not to provide specific forecasts for a given experiment, but rather to compare our method with a standard tomographic analysis, with both methods using the linear matter power spectrum and the same survey specifications.  The inclusion of nonlinearities does not add relevant information and can be disregarded without loss of generality.

\section{Galaxy number counts}\label{sec:methodology}
Firstly, we briefly describe two of the main approaches to galaxy clustering data: the  Fourier power spectrum, usually employed with spectroscopic data; and the 2D tomographic angular power spectrum, usually used in photometric surveys. Secondly, we introduce the new hybrid approach towards optimal angular power spectra for spectroscopic galaxy surveys.

\subsection{Fourier power spectrum}\label{ssec:3D}
Neglecting RSD and the Alcock-Paczynski (AP) effect for simplicity, the galaxy number density at real-space position ${\bm x}$ is 
\be
n_{\rm g}({\bm x})=\bar n_{\rm g}(z)\left[1+\delta_{\rm g}({\bm x},z)\right],
\ee
where $\bar n_{\rm g}(z)$ is the mean galaxy number density at redshift $z$ and the fluctuations $\delta_g$ have zero mean and same-time power spectrum $\big\langle\delta_{\rm g}({\bm k},z)\delta_{\rm g}^\ast({\bm k}^\prime,z)\big\rangle=(2\pi)^3\delta_{\rm D}({\bm k}-{\bm k}^\prime)P_{\rm g}(k,z)$, where $P_{\rm g}(k,z)=b_{\rm g}^2(z)P_\delta(k,z)$, $b_{\rm g}(z)$ is the galaxy bias and $P_\delta(k,z)$ the linear matter power spectrum, which is related to the (dimensionless) primordial power spectrum of curvature fluctuations, $\Delta^2_\zeta(k)=A_{\rm s}(k/k_0)^{\ns-1}$, through
\begin{equation}
P_\delta(k,z)=\frac{8\pi^2}{25}\frac{k_0A_{\rm s}}{\ho^4\om^2g_\infty^2}D_+^2(z)T_{\rm m}^2(k)\left(\frac{k}{k_0}\right)^\ns.
\end{equation}
Here $T_{\rm m}(k)$ is the matter transfer function (normalised so that $T_{\rm m}\to1$ for $k\to0$), $D_+(z)$ the growth factor of matter density fluctuations (normalised so that $D_+\to1$ for $z\to0$), and $g_\infty=\lim_{z\to\infty}(1+z)D_+(z)\approx 1.27$. The usual way of assessing the constraining power of a spectroscopic galaxy redshift survey is via the Fisher matrix, defined e.g.\ in \citet{Seo:2003pu}:
\begin{multline}
\mathbfss F\left(\vartheta_\alpha,\vartheta_\beta\right)=\sum_{i=1}^{N_{\rm b}}\frac{1}{4\pi^2}V(z_i)f_{\rm sky}\times\\
\int_{k_{\rm min}}^{k_{\rm max}}\de k\,k^2\frac{\partial P_{\rm g}(k,z_i)}{\partial\vartheta_\alpha}\frac{\partial P_{\rm g}(k,z_i)}{\partial\vartheta_\beta}
\left[P_{\rm g}(k,z_i)+\frac{1}{\bar n_i}\right]^{-2}.\label{eq:FisherPk}
\end{multline}
In this equation,  
$i$ labels the redshift bin centered in $z_i$, $N_{\rm b}$ is the number of redshift bins considered, $V(z_i)$ is the cosmological volume in bin $i$; $f_{\rm sky}$ is the fraction of sky surveyed; and $\bar n_i$ is the {\it volumetric} number density of galaxies in the  $i$th bin. In practice, in each redshift bin we count all the $k$-modes contained within the volume $V(z_i)f_{\rm sky}$, with an uncertainty accounting for both cosmic variance and the effect of discrete Poisson sampling.

The method relies on a number of assumptions, the most important being:
\begin{itemize}
\item[\textit{(i)}] All quantities are assumed constant within the redshift bin;
\item[\textit{(ii)}] No correlation among redshift bins is considered;
\item[\textit{(iii)}] It is valid only in the flat-sky limit;
\item[\textit{(iv)}] It does not include the effect of lensing.
\end{itemize}
A number of works in the recent literature have tried to overcome some of these limitations. For instance, \citet{Ruggeri:2017rza} apply a set of weights to extract RSD measurements as a function of redshift, acknowledging that future surveys covering a broad redshift range can no longer ignore cosmic evolution. \citet{Bailoni:2016ezz} improve upon the standard 3D $P(k)$ Fisher method by  taking into account three effects: the finite window function, the correlation between redshift bins and the uncertainty on redshift estimation. \citet{Gil-Marin:2015sqa} use a line-of-sight dependent power spectrum to deal with the large sky coverage of the Sloan Digital Sky Survey (SDSS) DR12 catalogue. Similarly, \citet{Blake:2018swt} study power spectrum multipoles in a curved sky for the 6-degree Field Galaxy Survey (6dFGS), and indicate how these techniques may be extended to studies of overlapping galaxy populations via multipole cross-power spectra. However, to our knowledge none of these methods includes lensing and none simultaneously addresses all of the three other issues.

\citet{Tansella:2017rpi} derive an exact expression for the galaxy correlation function in redshift shells---including lensing---thus addressing all four issues listed above. Their approach is an alternative to ours and is based on the  correlation function. However, they do not give forecasts to compare their method with previous work.

\subsection{2D tomographic angular power spectrum}\label{ssec:2D}
Uncertainties in photometric measurements prevent us from mapping angles $\hat{\bm n}$ and redshifts to three-dimensional positions $r\big(\chi(z),\hat{\bm n}\big)$, which we would, in turn, translate into ${\bm k}_\perp$ and $k_{||}$. Instead, we have to rely on angular power spectra, where the radial information is averaged over a redshift bin, taking  into account probability density functions of photometric redshifts. This leads to a tomographic matrix with entries
\begin{equation}
C^{ij}_\ell=4\pi\int\!\!\de\ln k\,\mathcal W^i_\ell(k)\mathcal W^j_\ell(k)\Delta^2_\zeta(k),\label{eq:Cl}
\end{equation}
where $\mathcal W^i_\ell(k)$ is the transfer function of the $i$-bin, containing weights for all the terms included, i.e., Newtonian density fluctuations,  RSD, lensing and possibly other relativistic contributions \citep[see e.g.][]{Camera:2014bwa}. Weak lensing convergence, modulated by the magnification bias, gives the strongest relativistic correction  \citep{Raccanelli:2015vla,Fonseca:2015laa}. We thus limit our analysis to this effect only, beyond density fluctuations and RSD.

The angular power spectrum transfer functions of Eq.~\eqref{eq:Cl} can be rewritten as
\begin{equation}
\mathcal W^i_\ell(k)=\int\!\!\de\chi\,\frac{\de N_{\mathrm g}^i}{\de\chi}\mathcal W_\ell(k,\chi).
\end{equation}
where we remind the reader that $(\de N_{\mathrm g}^i/\de\chi)\de\chi=(\de N_{\mathrm g}^i/\de z)\de z$. In longitudinal gauge,
\begin{align}
\mathcal W_\ell(k,\chi)=&\,b_{\rm g}(\chi){\delta}_k(\chi) j_\ell(k\chi)+\frac{k}{\mathcal H(\chi)}v_k(\chi)j^{\prime\prime}_\ell(k\chi) \nonumber\\&+2\left[\mathcal Q(\chi)-1\right]\ell(\ell+1)\kappa(\chi).\label{eq:den+RSD+len}
\end{align}
The first line is the main contribution, from density perturbations ($\delta$ is the comoving matter density contrast, necessary for a physical model of bias\footnote{The correction term from using the comoving $\delta$ is omitted since it is negligible on sub-Hubble scales.}) and RSD ($v$ is the peculiar velocity), and the second line is the lensing contribution, proportional to the weak lensing convergence
\begin{equation}
\kappa(\chi) = \frac{1}{2}
\int_0^\chi\!\!\de\tilde\chi\,\frac{(\chi-\tilde\chi)}{\chi\tilde\chi}\left[\Phi_k(\tilde\chi)+\Psi_k(\tilde\chi)\right]j_\ell(k\tilde\chi),
\end{equation}
with $\Phi$ and $\Psi$ the two gauge invariant Bardeen potentials.

For the tomographic case, the Fisher matrix takes the form (summation over same indexes is assumed)
\begin{equation}
\mathbfss F\left(\vartheta_\alpha,\vartheta_\beta\right)=\sum_{\ell=\lm}^{\lM}\frac{2\ell+1}{2}f_{\rm sky}\frac{\partial C^{i j}_\ell}{\partial\vartheta_\alpha}\left(\widetilde{\mathbfss C}_\ell\right)^{-1}_{jm}\frac{\partial C^{m n}_\ell}{\partial\vartheta_\beta}\left(\widetilde{\mathbfss C}_\ell\right)^{-1}_{ni},\label{eq:FisherCl}
\end{equation}
where $(\cdot)^{-1}$ denotes matrix inversion,  
\be
\widetilde C^{j m}_\ell=C^{j m}_\ell+\delta^{jm}\frac{1}{\tilde n_j}, \label{eq:Cl_obs}
\ee
is the $j$-$m$ entry of the observed tomographic matrix $\widetilde{\mathbfss C}_\ell$, and $\tilde n_j$ is the \textit{angular} number density of galaxies per steradian in bin $j$.

\subsection{Hybrid angular power spectrum}\label{ssec:2.5D}
In principle, the correct way to analyse spectroscopic galaxy clustering is via the  angular power spectra $C^{ij}_\ell$, including all cross-bin correlations. These power spectra are a physical and gauge-independent representation of correlations in galaxy number counts, while the Fourier power spectrum is not an observable in itself, since it is gauge dependent \citep[e.g.][]{Bonvin:2011bg}. In practice, however, this is computationally unfeasible, especially since many realisations are required  for simulated data. Our new method for computing angular power spectra for spectroscopically selected galaxies does not rely on the computation of an extremely large number of auto- and cross-correlations between redshift bins, but is still capable of constraining cosmological parameters with an accuracy comparable to that of a Fourier $P_{\rm g}({\bm k},z)$ analysis (see \S~\ref{ssec:constraints}). 

For the  $P_{\rm g}({\bm k},z)$ approach in \S~\ref{ssec:3D} (including RSD and AP corrections), one usually takes a rather thick redshift bin---with width $\sim\Delta z=0.1$---and then counts all 3D $k$ modes within the bin, disregarding correlations among all other redshift bins. This effectively renders the covariance matrix diagonal in redshift. On the other hand, if we were to do the same with the 2D tomographic approach of \S~\ref{ssec:2D}, we would lose substantial information by squashing all the galaxies contained within the $\Delta z$ bin onto a single redshift slice.

This motivates a hybrid approach: divide the redshift distribution into $\Delta z=0.1$ bins as in the $P_{\rm g}({\bm k},z)$ approach; subdivide each bin into 10 top-hat bins with width $\delta z=0.01$; convolve the thin bins with a Gaussian with spread $\sigma_{\rm z}=0.001$ to account for the small but non-negligible errors in the spectroscopic redshift estimation.

\begin{figure}
\centering
\includegraphics[width=\columnwidth]{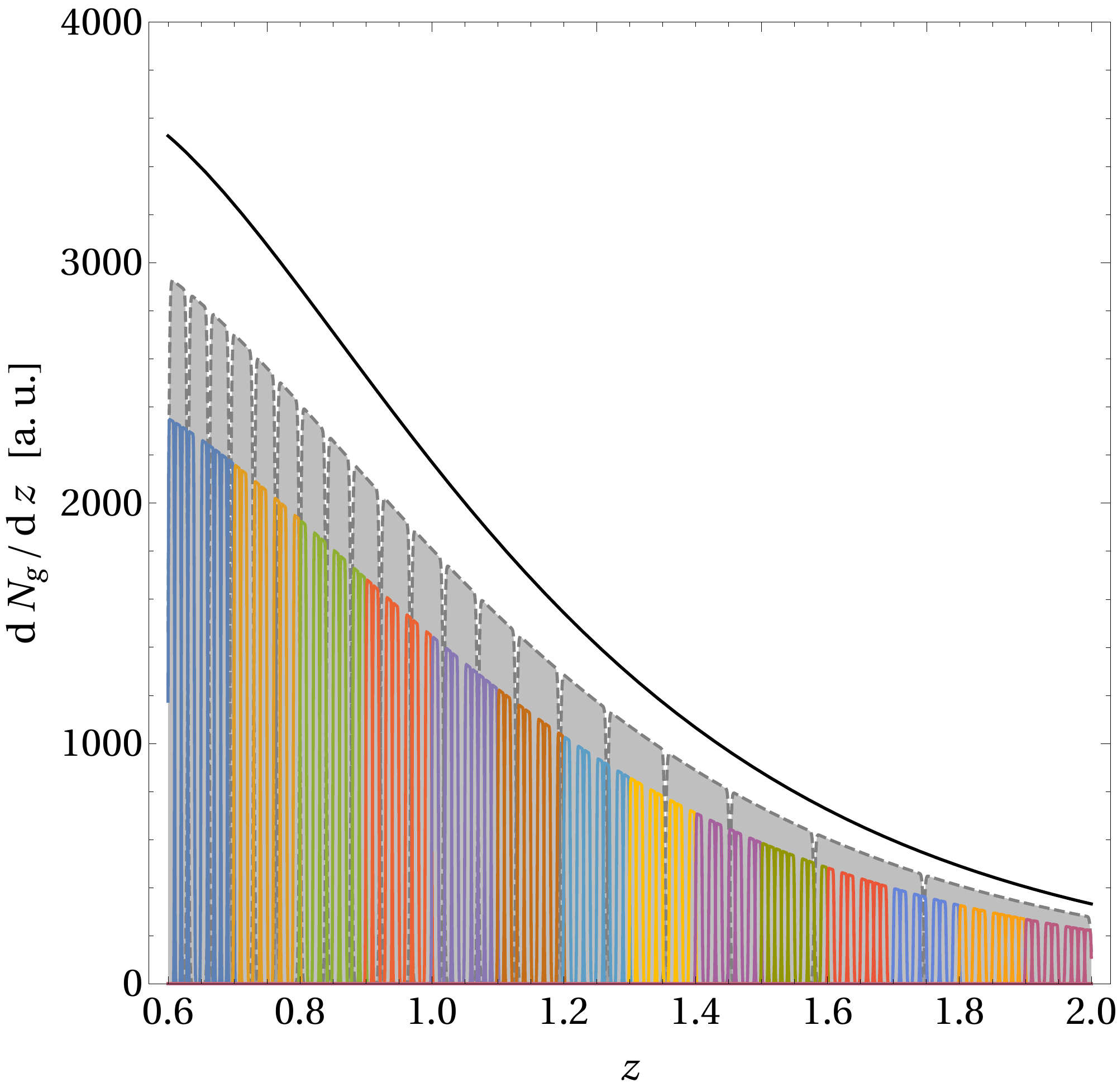}
\caption{Estimated redshift distribution of spectroscopically selected H$\alpha$ galaxies with flux $>3\times10^{-16}\,\mathrm{erg\,cm^{-2}\,s^{-1}}$ \protect{\citep[from][]{Pozzetti:2016cch}}. The top  black curve shows the unbinned $n_{\rm g}(z)$, as in Eq.~\eqref{eq:ng}. The shaded grey regions bounded by dashed lines are the 20 equi-populated  bins of a standard tomographic analysis (\S~\ref{ssec:2D}). The bundles of coloured curves depict the 14 equi-spaced coarse redshift bins with width $\Delta z=0.1$, in turn subdivided into 10  thinner top-hat bins of width $\delta z=0.01$, finally convolved with  a Gaussian ($\sigma_{\rm z}=0.001$), as in \S~\ref{ssec:2.5D}. (Note that all distributions have been rescaled by an arbitrary factor to enhance readability.)}\label{fig:n_g-2.5D}
\end{figure}

The set-up is illustrated in Fig.~\ref{fig:n_g-2.5D}, where:
\begin{itemize}
\item[\textit{(i)}] We start from the redshift distribution in Eq.~\eqref{eq:ng} of galaxies expected to be detected by a Stage IV cosmological survey with H$\alpha$ flux $>3\times10^{-16}\,\mathrm{erg\,cm^{-2}\,s^{-1}}$ \citep[upper, black curve, as in Model 3 of][]{Pozzetti:2016cch};
\item[\textit{(ii)}] We first bin galaxies into 14 equi-spaced, coarse redshift bins (groups of curves with the same colour);
\item[\textit{(iii)}] We subdivide the coarse bins into 10 thinner top-hat bins;
\item[\textit{(iv)}] We finally convolve the thinner bins with the $\sigma_{\rm z}=0.001$ Gaussian (a total of 140 distinct, slightly overlapping curves).
\end{itemize}

We effectively consider each thick bin as a separate survey, using the full spectroscopic angular power spectra in each thick bin, by applying Eq.~\eqref{eq:FisherCl} with respect to its sub-bins. Then, as in Eq.~\eqref{eq:FisherPk}, we sum the 14 Fisher matrices thus obtained. This hybrid approach stems from the more physically motivated angular power spectra, but resembles what is done with the $P_{\rm g}({\bm k},z)$ analysis, since it preserves radial information within the thick redshift bins by considering all the cross-correlations between the thin sub-bins. 

The hybrid method is far less computationally onerous than the full 3D angular power spectra, which involve application of Eq.~\eqref{eq:FisherCl} to a spectroscopic tomographic matrix with bin width 0.01. For $0.6\le z\le2$, this involves a total of $9870$ spectra, between auto- and cross-bin correlations. By contrast, our hybrid method only requires $770$ of them, more than one order of magnitude fewer.

\begin{figure*}
\centering
\includegraphics[width=0.45\textwidth]{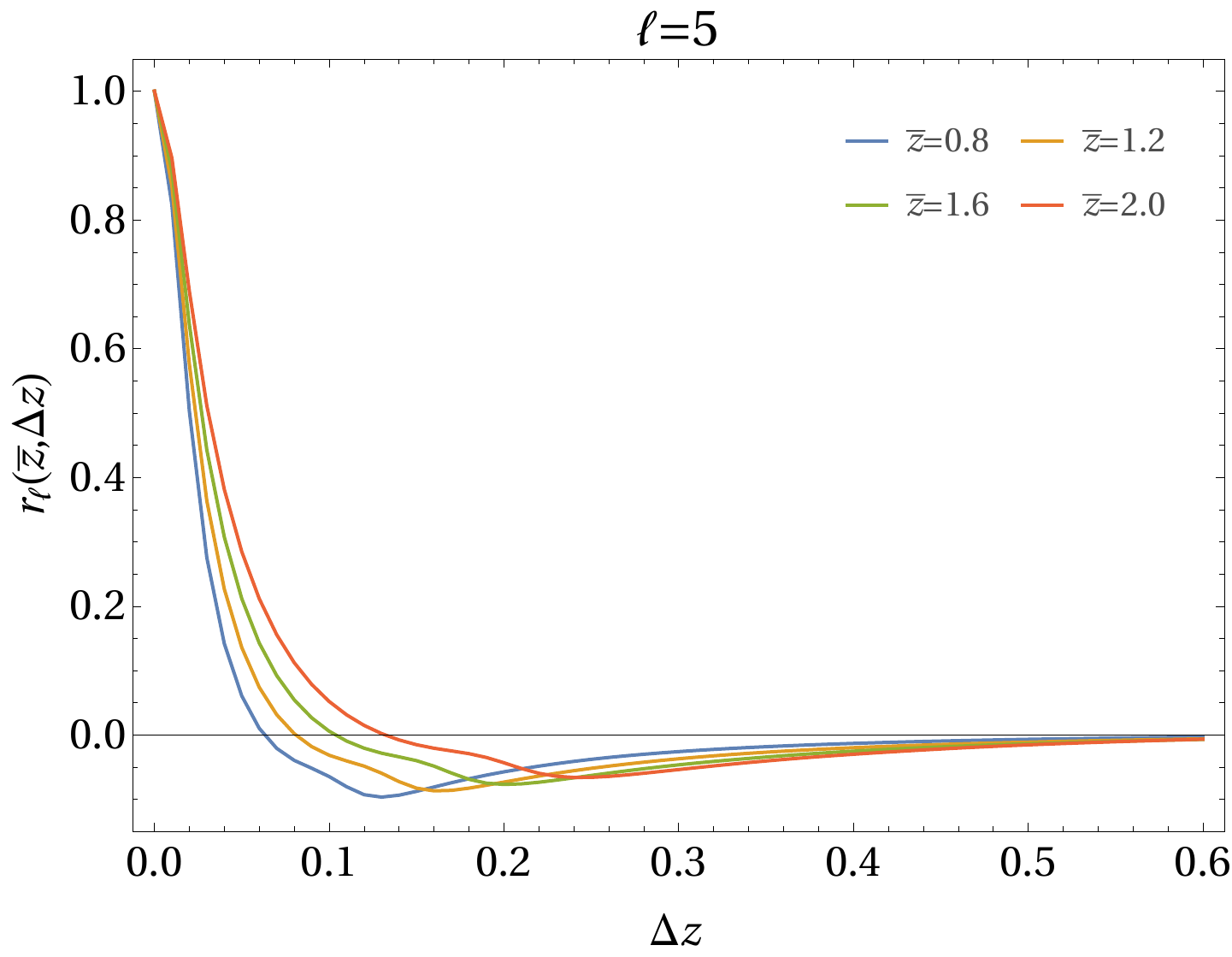}
\hspace{0.05\textwidth}\includegraphics[width=0.45\textwidth]{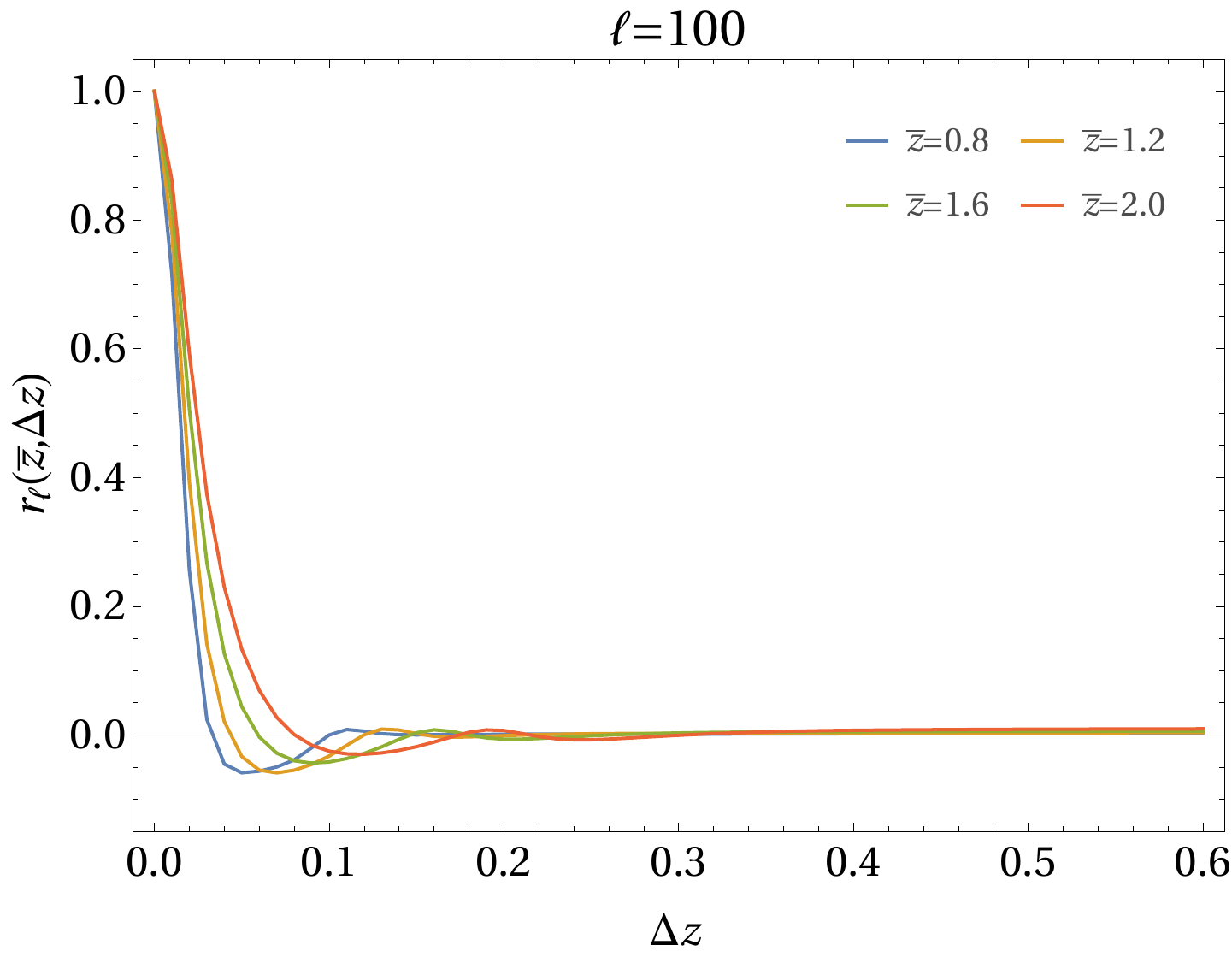}
\caption{Correlation coefficient $r_\ell(\bar z,\Delta z)$ as a function of redshift separation between bins centred in $\bar z = 0.8,1.2,1.6$ and $2.0$, for two reference multipoles, $\ell=5$ and $\ell=100$.}\label{fig:r_vs_dz}
\end{figure*}

In principle, by considering each thick bin as an independent survey, we could neglect valuable information encoded in cross-correlations among distant redshift bins, or between thin bins at the edges of the thick bins. This is what is usually done in observational $P_{\rm g}({\bm k},z)$ analyses, but it is not clear how one can assess the loss of information in that case \citep[for a possible solution, see e.g.][]{Bailoni:2016ezz}. In our case, we can assess the effects of neglecting such cross-bin correlations by looking at the correlation coefficient between bins. If we consider $C_\ell(z_1,z_2)$, the angular cross-power spectrum of galaxy number counts between redshifts $z_1$ and $z_2$, integrated over a thin slice of width $\delta z=0.01$, we can define the correlation coefficient as
\be
r_\ell(z_1,z_2)\equiv \frac{C_\ell(z_1,z_2)}{\sqrt{C_\ell(z_1,z_1)C_\ell(z_2,z_2)}},\label{eq:r_l}
\ee
where we make explicit the dependence on the angular scale, $\ell$. Alternatively, we can redefine $r_\ell(\bar z,\Delta z)$ for two bins centered in $\bar z$ and separated by $\Delta z$, with $z_1$ and $z_2$ of Eq.~\eqref{eq:r_l} thus becoming $\bar z-\Delta z/2$ and $\bar z+\Delta z/2$, respectively. We emphasize that here $\Delta z$ does not refer to the width of the thick bins introduced above, but we nevertheless employ this slight abuse of notation for a reason that will become clear when looking at Fig.~\ref{fig:r_vs_dz}. There, we plot $r_\ell(\bar z,\Delta z)$ as a function of the redshift separation between bins of width $\delta z=0.01$, centered in redshifts $\bar z = 0.8,1.2,1.6$ and $2.0$, for angular scales $\ell=5$ and $100$. It is clear that the correlation coefficient quickly falls as a function of $\Delta z$, thus indicating that most of the information comes from correlations at small radial separation.

The take-home message is as follows. We have assumed that all correlations carry similar information about the parameters we are interested in, and treated each thick $\Delta z =0.1$ slice as an independent survey. Figure~\ref{fig:r_vs_dz} shows that this a reasonable assumption. Nevertheless, we shall come back later to this for further checks (see \S~\ref{ssec:tests}).

\section{Results and discussion}\label{sec:results}
\subsection{Constraints on cosmological parameters}\label{ssec:constraints}
First of all, we focus on the constraining power of the hybrid approach on \lcdm\ cosmological parameters. As explained above, we limit our analysis to linear scales and set the maximum multipole for $C^{i j}_\ell$ to
\begin{equation}
\lM(z_i,z_j)=\min\left[\chi(z_i),\chi(z_j)\right]k_{\rm nl}.
\end{equation}
The maximum angular scale is determined by $4\pi f_{\rm sky}$, i.e.\ $ 15,000\,{\rm deg}^2$. Thus $\lm=2$.

\begin{figure}
\centering
\includegraphics[width=\columnwidth]{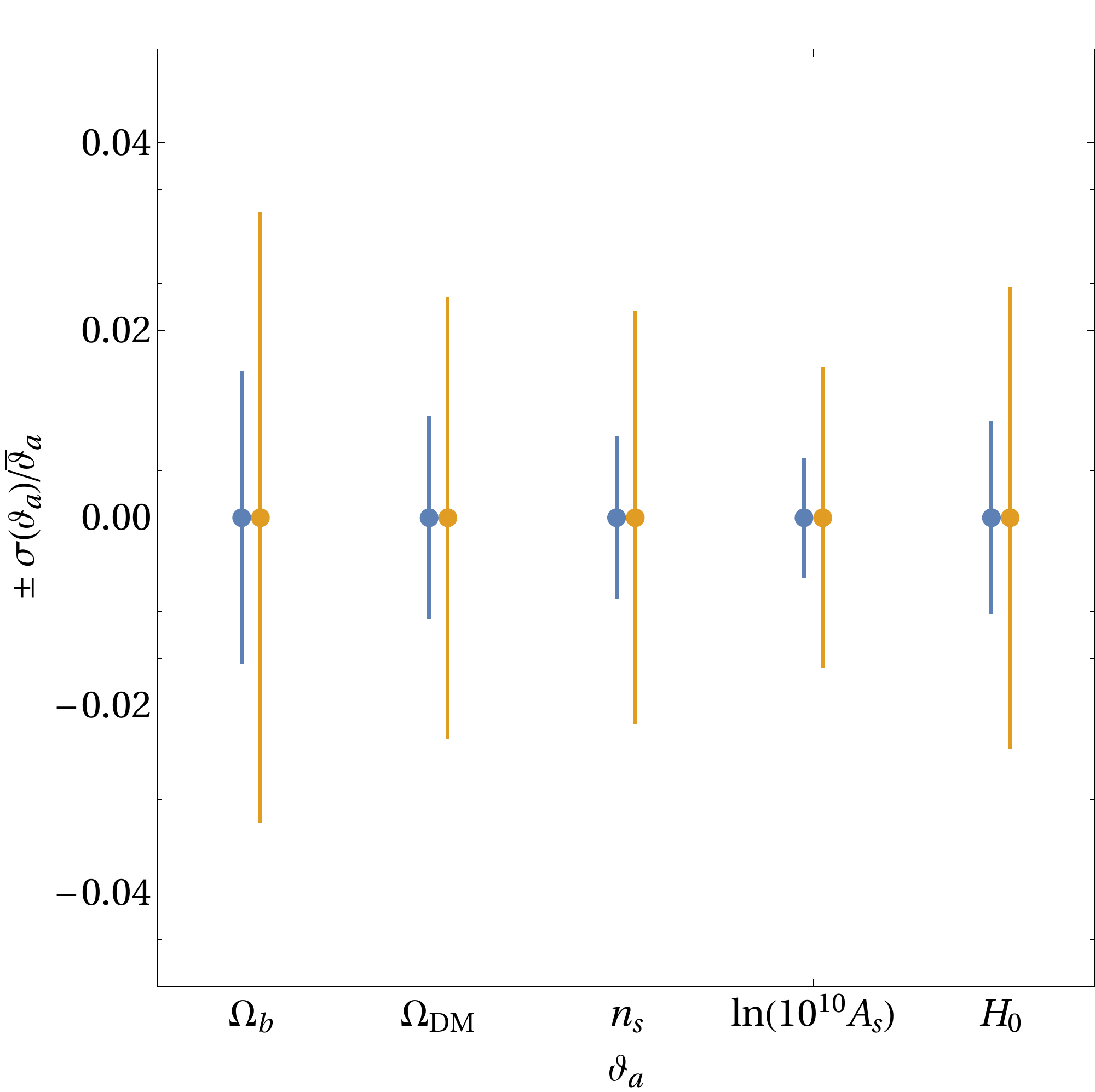}
\caption{Relative errors on cosmological parameters obtained with our hybrid binning (blue) compared to standard tomography (yellow).}\label{fig:LinearErrors}
\end{figure}

In Fig.~\ref{fig:LinearErrors}, we compare the constraining power of our hybrid approach (blue error bars) to that of a standard tomographic analysis (yellow error bars). Each cosmological parameter $\vartheta_\alpha$ is rescaled by its fiducial value $\overline{\vartheta}_\alpha$, to focus more easily on the relative tightness of the forecast 1$\sigma$ constraints. It is clear that the finer binning of our hybrid method retrieves more information from the spectroscopic galaxy clustering data even on strictly linear scales, where for instance the largest angular multipole allowed is $\lM(z_{140},z_{140})=718$.

So far, we have focused on the cosmological parameters, disregarding any uncertainty on survey specifics such as the galaxy bias. Now we introduce nuisance parameters by adding one parameter per redshift bin, thus allowing for a freely-varying bias amplitude in each bin. We emphasise that this is a conservative approach, as it does not make use of any prior knowledge on the bias. As a result, the standard 2D tomographic approach has 20 nuisance parameters, which are in turn marginalised over (see \S~\ref{ssec:marginalisation}). For the hybrid approach, the situation is slightly different. As each coarse redshift bin is further subdivided into 10 thinner bins, a total of 140 nuisance parameters is included. However, each $\Delta z=0.1$ thick bin is effectively treated as independent from the others, as in the $P_{\rm g}({\bm k},z)$ case of Eq.~\eqref{eq:FisherPk}. Thus, marginalisation over nuisance parameters is performed separately for each Fisher matrix pertaining to a specific thick bin, and only at the end are all the marginal Fisher matrices summed up to give the final cosmological parameter Fisher matrix.For simplicity we fix  the smallest angular scale to a nominal $\lM=800$.

Table~\ref{tab:cosmo_constraints} shows the resulting $1\sigma$ relative marginal errors on \lcdm\ cosmological parameters. We find that our constraints are comparable to those from a standard $P_{\rm g}({\bm k},z)$ analysis (including AP corrections): this can be seen by comparing the last row of Table~\ref{tab:cosmo_constraints} with the 4th column of Table~2 in \citet{Bailoni:2016ezz}, who use \textit{Euclid} specifications that are the same as those in \S~\ref{obas}, with the same number of thick bins. The first row of Table~\ref{tab:cosmo_constraints} shows constraints from standard broad-bin tomography. The hybrid approach is approximately twice as constraining as a standard tomographic binning, despite the larger number of nuisance parameters.
\begin{table*}
\centering
\caption{Forecast 1$\sigma$ relative errors  on cosmological parameters, after marginalising over nuisance bias parameters, for the two approaches.}
\begin{tabular}{lccccc}
$(\lm,\lM)=(2,800)$& $\sigma(\ob)/\overline{\Omega}_{\rm b}$ & $\sigma(\odm)/\overline{\Omega}_{\rm DM}$ & $\sigma(\ns)/\overline{n}_{\rm s}$ & $\sigma(\As)/\overline{\mathcal A}_{\rm s}$ & $\sigma(\ho)/\overline{H}_0$\\
\hline
Standard tomography (\S~\ref{ssec:2D}) & $4.8\%$ & $3.3\%$ & $2.5\%$ & $2.2\%$ & $3.1\%$ \\
Hybrid tomography (\S~\ref{ssec:2.5D}) & $2.5\%$ & $1.7\%$ & $1.1\%$ & $0.9\%$ & $1.6\%$
\end{tabular}\label{tab:cosmo_constraints}
\end{table*}
\begin{table*}
\centering
\caption{Forecast relative biases on cosmological parameters,  after marginalising over nuisance bias parameters, for the two approaches.}
\begin{tabular}{lccccc}
$(\lm,\lM)=(2,800)$& $b(\ob)/\sigma(\ob)$ & $b(\odm)/\sigma(\odm)$ & $b(\ns)/\sigma(\ns)$ & $b(\As)/\sigma(\As)$ & $b(\ho)/\sigma(\ho)$\\
\hline
Standard tomography (\S~\ref{ssec:2D}) & $8.4$ & $11.1$ & $-8.9$ & $0.2$ & $6.0$ \\
Hybrid tomography (\S~\ref{ssec:2.5D}) & $0.11$ & $-0.22$ & $0.08$ & $0.05$ & $-0.10$
\end{tabular}\label{tab:cosmo_bias}
\end{table*}

\subsection{Bias on parameter estimation from neglecting lensing}\label{ssec:biases}
As discussed in \S~\ref{ssec:3D}, future spectroscopic galaxy surveys will cover an unprecedented redshift range, so that the effect of  weak lensing convergence on number counts needs to be assessed. This is straightforward to do with angular power spectra, which  naturally include the weak lensing contribution. This contribution is computationally intensive, and here we evaluate the consequences of ignoring it in the interests of speeding up computations.

Following \citet{Camera:2016owj} and \citet{Fonseca:2015laa}, we introduce a `fudge' factor parameterising the \textit{theoretical systematic effect} represented by neglecting the lensing contribution in galaxy count angular power spectra. Then Eq.~\eqref{eq:den+RSD+len} becomes
\be
\mathcal W_\ell=
b_{\rm g}{\delta}_k\, j_\ell
+\frac{k}{\mathcal H}{v}_k\,j^{\prime\prime}_\ell+2A_\kappa\big(\mathcal Q-1\big)\ell(\ell+1)\kappa,
\ee
where the parameter $A_\kappa$ has fiducial value $\overline{A}_\kappa=1$ when lensing is correctly included in the analysis, and $A_\kappa=0$ when lensing is neglected. Note that intermediate values of $A_\kappa$ have a physical meaning if we wish to account for an uncertainty in the magnification bias term $(\mathcal Q-1)$, which modulates the contribution from lensing convergence.

The Fisher matrix  technique for computing bias in parameters is outlined in  Appendix~\ref{ssec:bias}. In the present case, the vector of shifts $\delta\varphi_a$ reduces to $\delta A_\kappa\equiv(\overline{A}_\kappa-0)=1$. Table~\ref{tab:cosmo_bias} shows the bias on cosmological parameters in units of standard deviations, for a standard 2D tomographic analysis (first row) and when implementing our hybrid approach (second row). As expected, standard tomography is very sensitive to lensing effects, because its wider redshift bins acquire a significant contribution from the integrated effect of lensing. On the other hand, the thinner the bin, the less important is lensing \citep[see also][]{Villa:2017yfg}. Thus, our hybrid approach, with its thin sub-bins, is almost insensitive to it, and we can therefore safely ignore the lensing contribution in the hybrid approach. A major advantage is that this allows for a significant speed-up of computations.

\subsection{Information gain}\label{ssec:IG}
An alternative way to assess the enhancement in constraining power delivered by the hybrid approach over standard 2D tomography is a statistical tool called `information gain'. When we compare the parameter posterior distributions from two experiments, $p_1(\boldsymbol\vartheta)$ and $p_2(\boldsymbol\vartheta)$, the information gain in going from $p_1$ to $p_2$ is $D(p_2||p_1)$. Information gain (also known as `relative entropy') was originally motivated by information theory, but it can be used to compute the information gained by Bayesian updates in units of bits (see Appendix~\ref{sapp:IG}).

One of the most useful properties of information gain is that it is invariant under invertible transformations in the random variable ${\bm x}$. In other words, it represents a more agnostic way to compare two experimental set-ups, because re-parametrisations of the parameter set $\boldsymbol\vartheta\equiv\{\vartheta_\alpha\}$ do not affect the information gain. This frees us from misleading interpretations of forecast parameter constraints, which can look tighter or looser depending on the choice of the parameter basis.

For two alternative ways to analyse  spectroscopic galaxy surveys, the correct way to use information gain is to compare the gain the two methods (i.e.\ 2D tomography and our hybrid approach) have over a common prior information on cosmological parameters \citep{Grandis:2015qaa}. Namely, if $p_\Pi(\boldsymbol\vartheta)$ is the prior of the parameter set, we compute $D(p_{\rm 2D}||p_\Pi)$ and $D(p_{\rm hyb}||p_\Pi)$ separately, and we then compare the incremental gain.

We adopt Gaussian priors on the cosmological parameter set described by a Fisher matrix $\mathbfss F_\Pi$, whose entries are $\sigma^{-2}(\boldsymbol\vartheta)$. Following \citet{Raveri:2016xof}, we take $\sigma^{-2}(\{\ob h^2,\odm h^2,h,\As,\ns,\tau\})=\{100,1,6.25,0.25,0.1\}$. By using Eq.~\eqref{eq:IG_Gaussian} for both standard tomography and our hybrid approach on the same prior $p_\Pi$, we obtain an increment in information gain of $5.25\,{\rm bits}$ in favour of the hybrid method. According to \citet[][and references therein]{Grandis:2015qaa}, this is similar to the increment on WMAP9 represented by a compilation of BAO data from the 6dFGS, BOSS in SDSS III, WiggleZ and SDSS DR7. As a comparison, the increment brought by \planck\ data is $10\,{\rm bits}$, whereas weak lensing from CFHTLenS gives $1.7\,{\rm bits}$.

\subsection{The case of primordial non-Gaussianty}\label{ssec:PNG}
Detection of nonzero values of the primordial non-Gaussianity parameter $\fnl$ is one of the main goals of forthcoming galaxy surveys \citep[e.g][]{Amendola:2012ys,Amendola:2016saw,Camera:2015fsa,Alonso:2015uua}. It is a difficult measurement, since the strongest signal arises from scale-dependent galaxy bias on ultra-large scales. This is in sharp contrast to the standard cosmological parameters, which rely on the highest possible number of available modes---thus implying a strong preference towards small scales. Our hybrid approach is geared towards the standard parameters and we do not expect it to perform optimally on $\fnl$.

To quantify this, we now include $\fnl$ in the set of cosmological parameters, restricting the remaining parameters to those that are most degenerate with it, namely the amplitude of primordial fluctuations, $\As$, and the dark matter fraction, $\odm$. As before, we include a free amplitude in each redshift bin as a nuisance parameter, then marginalise over all parameters but $\fnl$. 

We find that the forecast marginal errors on the $\fnl$ increase by a factor of $\sim2$ when moving from standard 2D tomography to our hybrid  approach. On the other hand, the bias on the estimation of $\fnl$ induced by neglecting the lensing term in Eq.~\eqref{eq:den+RSD+len} is smaller with our hybrid method, as summarised in Table~\ref{tab:cosmo_bias-fNL}.
\begin{table}
\centering
\caption{Forecast relative bias on the non-Gaussianity parameter, $\fnl$,  after marginalising over nuisance bias parameters and $\odm$ and $\As$, for the two approaches.}
\begin{tabular}{lc}
$(\lm,\lM)=(2,800)$& $b(\fnl)/\sigma(\fnl)$\\
\hline
Standard tomography (\S~\ref{ssec:2D}) & $3.0$ \\
Hybrid tomography (\S~\ref{ssec:2.5D}) & $1.2$
\end{tabular}\label{tab:cosmo_bias-fNL}
\end{table}
The fact that both biases are $>1\sigma$ confirms the findings of \cite{Camera:2014sba} that it is necessary to include lensing---as well as other relativistic large-scale corrections---when trying to measure ultra-large scale effects such as primordial non-Gaussianity.

\subsection{Validity of the working assumptions}\label{ssec:tests}
To assess how much the neglect of correlations with distant thin bins impacts our results, we focus first on the thick bins. For each thick bin, we compute 10 Fisher matrices, starting from the full angular power spectrum tomographic matrix used so far, $C^{ij}_\ell$, and then removing off-diagonal elements in $C^{ij}_\ell$ as follows: the second Fisher matrix is computed from all $C^{ij}_\ell$'s except for the two entries at the extrema of the anti-diagonal, which are set to zero; and so on until we compute the 10th Fisher matrix, where $C^{ij}_\ell$ is nonzero only for $i=j$. 

\begin{figure}
\centering
\includegraphics[width=\columnwidth]{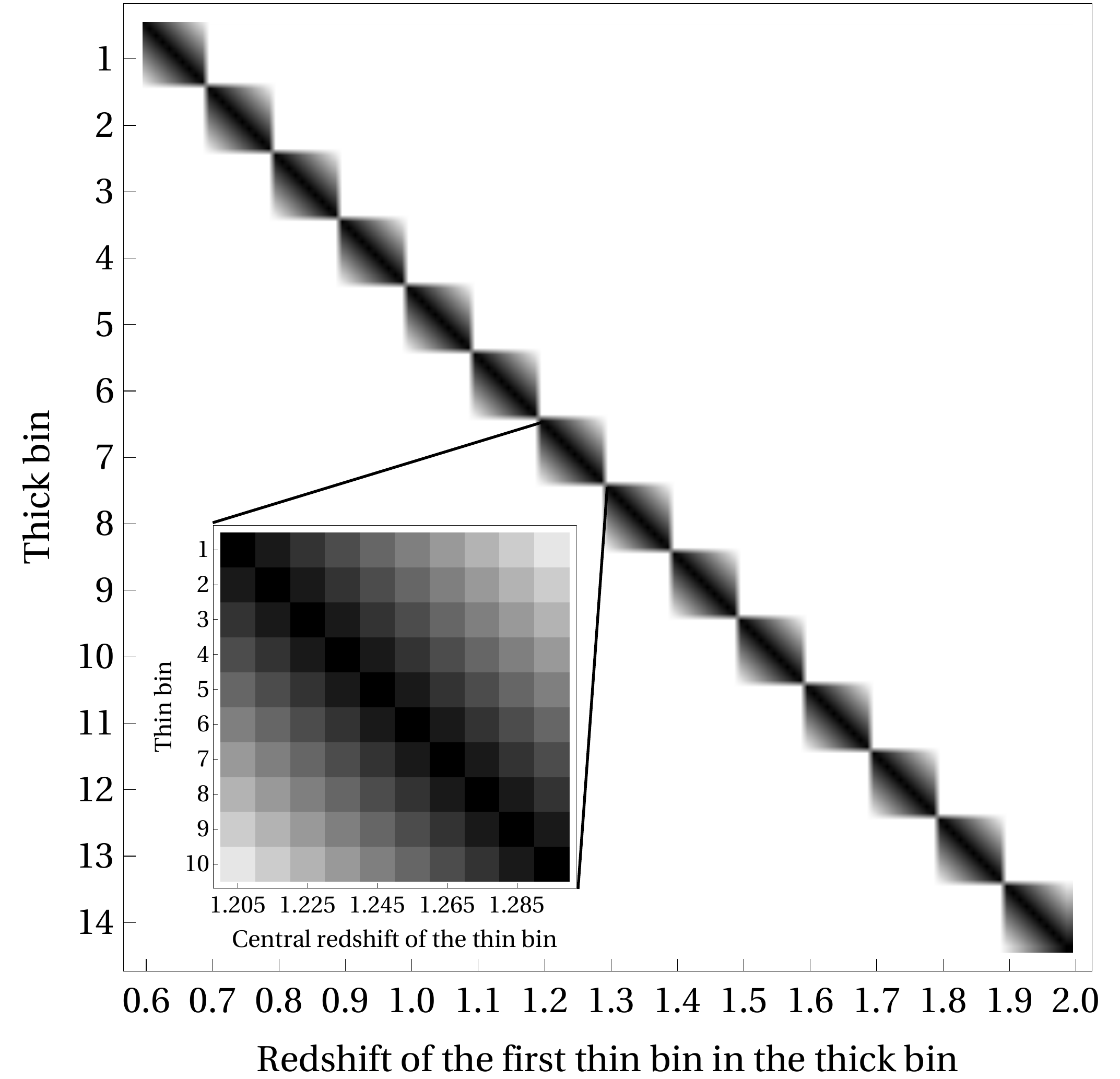}
\caption{General structure of the full angular power spectrum tomographic matrix, $C^{ij}_\ell$, in our hybrid approach. Block diagonal matrices represent the thick bins, which we consider as independent. The inset illustrates the structure of one of those thick bins, with the thin bins shown. Shades of grey refer to the tests described in \S~\ref{ssec:tests}.}\label{fig:structure}
\end{figure}

The inset in Fig.~\ref{fig:structure} illustrates this procedure, with shading going from black (only diagonal elements considered) to the lightest of grey (all the cross-correlations between thin bins considered). The larger figure shows the structure of the full tomographic matrix of our hybrid method for this test of the importance of off-diagonal correlations. In this case, white means no correlation included.

\begin{figure*}
\centering
\includegraphics[width=0.45\textwidth]{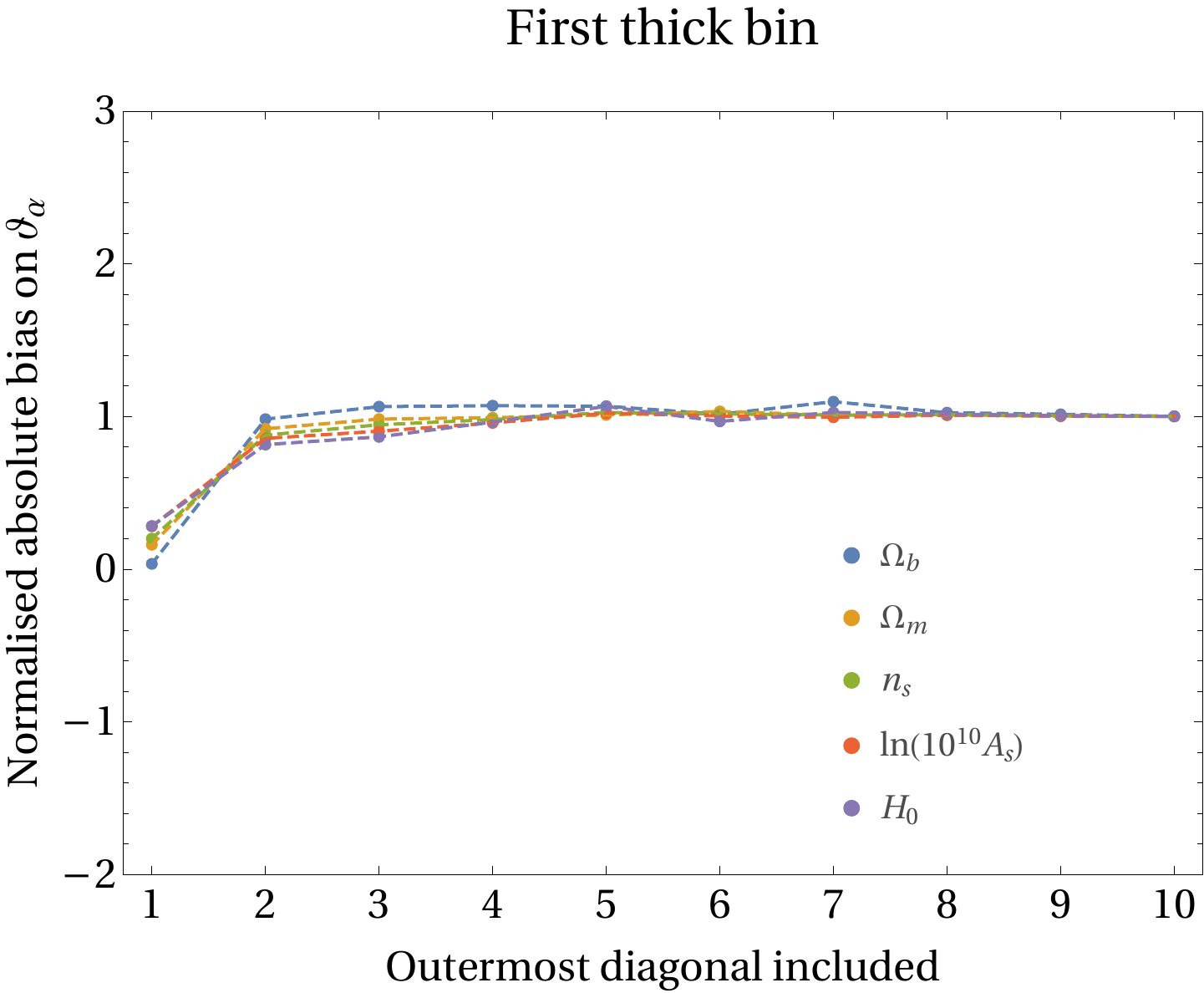}\hspace{0.05\textwidth}
\includegraphics[width=0.45\textwidth]{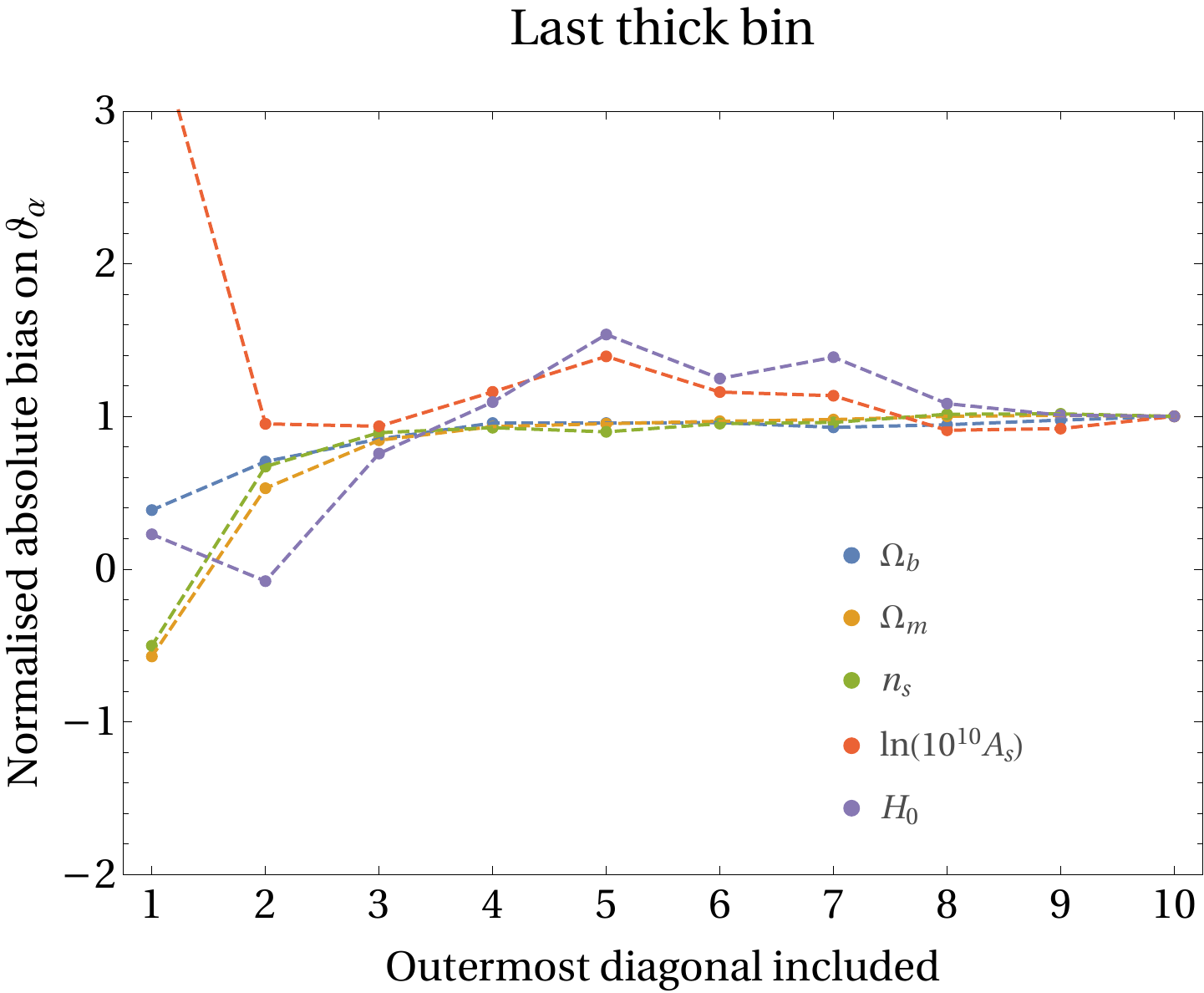}
\caption{Absolute bias on cosmological parameters induced by neglecting lensing magnification, as function a of off-diagonal elements included, for an analysis carried out either in the first or in the last of the thick redshift bins only, normalised to what is obtained by including all cross-correlations between thin bins in the given thick bin.}\label{fig:bias_thin}
\end{figure*}

Now, we consider two extrema: the first and the last of the thick bins. Referring to the inset of Fig.~\ref{fig:structure}, for both of these thick bins, we start from the full angular power spectrum tomographic matrix and at each step we remove one off-diagonal, from the lightest grey until we remain only with the black, main diagonal. In Fig.~\ref{fig:bias_thin} we illustrate the result of this test, by plotting the absolute bias on cosmological parameters induced by neglecting lensing magnification (normalised to the reference values of Table~\ref{tab:cosmo_bias}), as a function of how many cross-correlations between thin bins we include. Apart from the first point(s), for which the bias may be lower, but just because the Fisher matrices are very noisy, it appears clear that the behaviour is very flat. As expected, the trend is flatter at low redshift in the first thick bin. 

\begin{figure}
\centering
\includegraphics[width=\columnwidth]{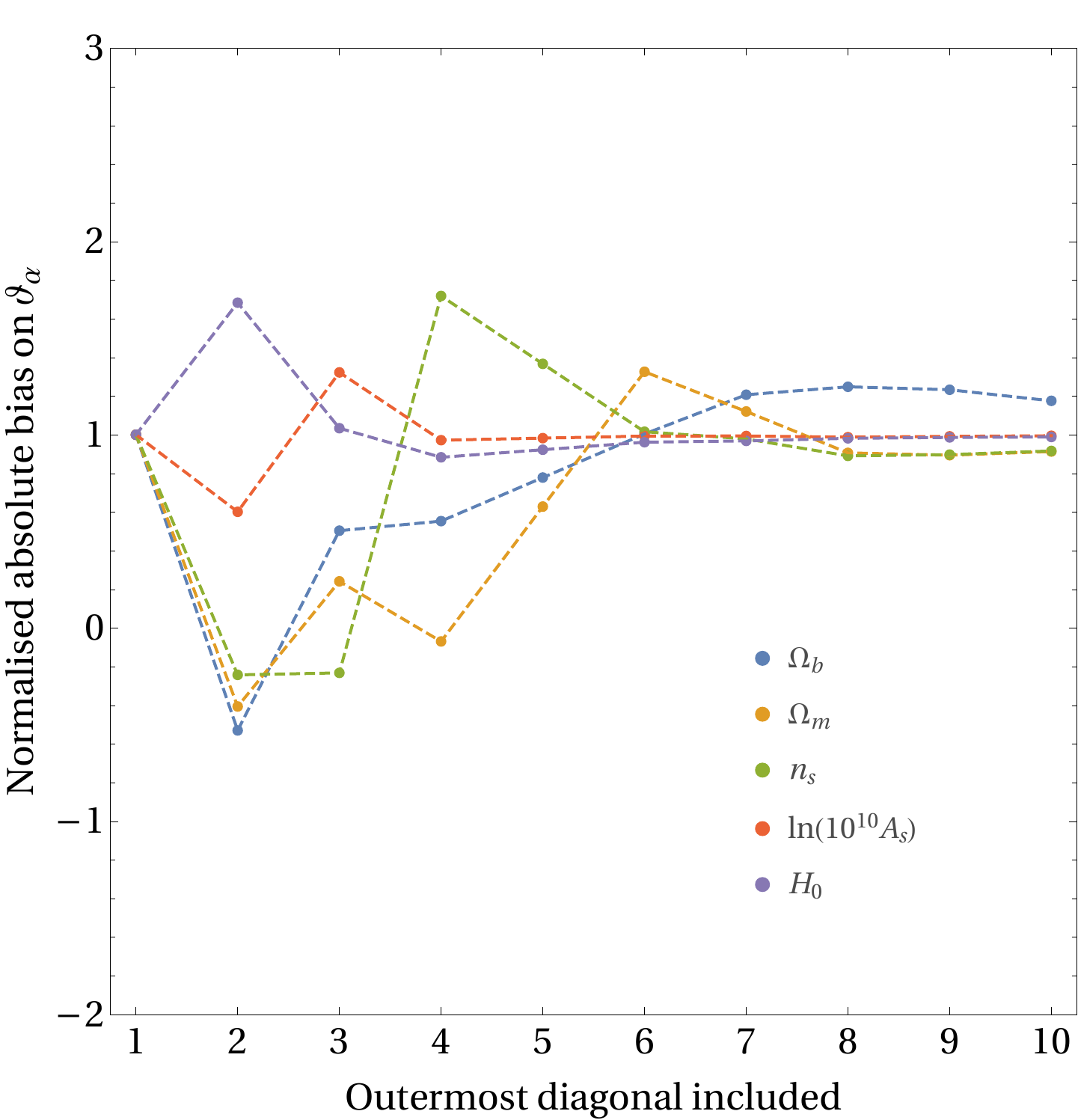}
\caption{The same as in Fig.~\ref{fig:bias_thin} but for the full angular power spectrum tomographic matrix of the hybrid method, namely the larger matrix in Fig.~\ref{fig:structure}. Here, each point is normalised to the previous one.}\label{fig:bias_full}
\end{figure}

Figure~\ref{fig:bias_full} shows the bias on cosmological parameters due to neglecting lensing, as a function of how many off-diagonal spectra are considered in the full angular power spectrum, i.e. the larger matrix in Fig.~\ref{fig:structure}. On the horizontal axis, 1 means black elements in the larger matrix, all the way to 10, representing the inclusion of all elements up to the lightest grey shade. Each point is normalised to its previous value, e.g.\ the 7th blue bullet point tells us that the bias on $\ob$ will increase by $\sim20\%$ if we include up to the 7th off-diagonal, with respect to what we would guess by considering only up to the 6th. We notice a net trend: the bias on cosmological parameter best-fit values induced by neglecting lensing may be significant if we considered only the main diagonal or the first few off-diagonals, but then it does not change any longer. This means that in our hybrid method the information encoded in the lensing term $\propto(\mathcal Q-1)\kappa$ flattens out more rapidly. This is mainly due to the thin slicing adopted for the sub-bins, in which lensing does not contribute substantially, as also noted by \citet{Villa:2017yfg}.

Finally, we perform a last test. After having checked that cross-correlations at separations larger that our chosen $\Delta z=0.1$ fall out rapidly (see Fig.~\ref{fig:r_vs_dz}), and that the bias on cosmological parameters due to neglecting lensing stabilises well before all the distant-bin correlations have been included (this section), we now investigate what happens to those thin redshift bins at the edges of the thick bins---where the blocks along the diagonal meet, in Fig.~\ref{fig:structure}. Those thin bins are nominally separated by less than $\Delta z$ but are not included in the analysis because they pertain to different thick bins, which are considered as separate surveys. To address this issue we proceed as follows:
\begin{itemize}
\item[\textit{(i)}] We recompute the Fisher matrix for the hybrid approach starting from the full tomographic $C^{ij}_\ell$ matrix as in Fig.~\ref{fig:structure}, rather than summing up all the Fisher matrices for each thick bin;
\item[\textit{(ii)}] We then add cross-correlations at the edges of the thick bins, basically as if we were overlapping more blocks on the main diagonal.
\end{itemize}
\begin{figure}
\centering
\includegraphics[width=\columnwidth]{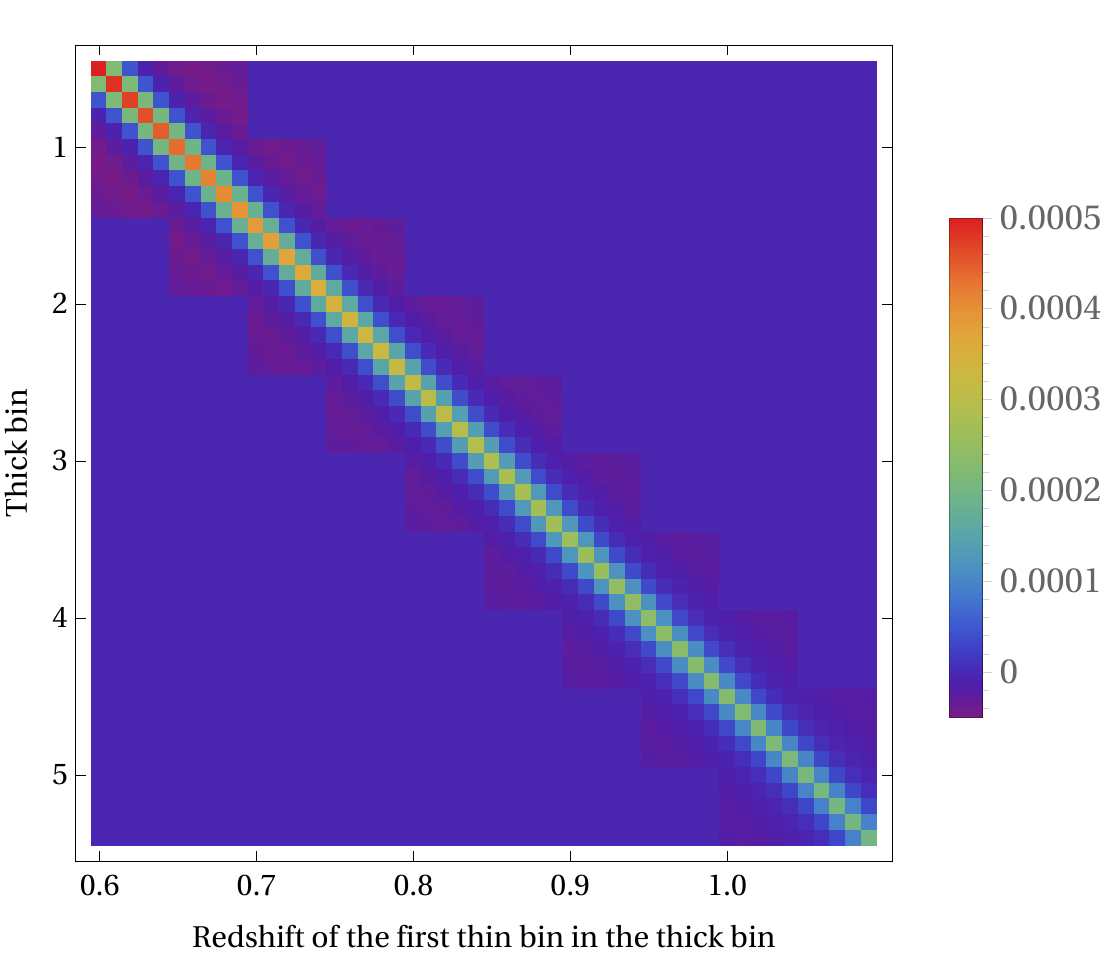}
\caption{Observed tomographic angular power spectrum, $\widetilde{\mathbfss C}_\ell$, at $\ell=2$ for the first five thick redshift bins.}\label{fig:covariance}
\end{figure}
The first computation $(i)$, by definition, has to give the same results as have been presented so far. Nonetheless, this is an important test for our code to pass, given the possible numerical instabilities in the inversion of a noisy $140\times140$ matrix. Then, we illustrate the second computation $(ii)$ in Fig.~\ref{fig:covariance}, where we show the observed angular power spectrum of Eq.~\eqref{eq:Cl_obs}---that is to say, signal plus noise---at $\ell=2$ for the first five thick bins, for the sake of clarity. As expected, some small non-zero correlation is present, although we emphasise that this is the very largest angular scale probed, and from Fig.~\ref{fig:r_vs_dz} we know that these correlations not only disappear rapidly as the separation in redshift increases, but also at larger multipoles.

Once we have constructed this upgraded tomographic matrix for the observed signal, we insert it into Eq.~\eqref{eq:FisherCl} and compute a new Fisher matrix, which now includes additional noise due to the off-diagonal correlations between thin bins at the edges of the thick bins. We find that forecast marginal errors change by less than $\sim2\%$ for all cosmological parameters. Let us emphasise that, by not including nuisance parameters, in the present case the sensitivity to cosmological parameters is the highest. Therefore, in a more realistic case, the contribution due to those correlations will be even more suppressed. This last check further confirms the validity of our assumption of using the thick bins as separate surveys.

\section{Conclusions}\label{sec:conclusions}
In this paper, we have presented a novel approach to optimise the use of angular power spectra with spectroscopic galaxy survey data. Our work is part of a community-wide effort aiming at crafting the best tools to exploit the oncoming wealth of cosmological data from large-scale structure experiments, such as the Square Kilometere Array, the European Space Agency's \textit{Euclid} satellite, or the Large Synoptic Survey Telescope. We emphasise that the approach outlined here is meant to be complementary to others \citep[see e.g.][]{Asorey:2012rd,Bailoni:2016ezz,Tansella:2017rpi}.

Our work is motivated by two facts. On the one hand, future galaxy surveys will reach unprecedented depths and sky coverages, even with spectroscopy. On the other hand, the techniques usually employed to estimate the 3D galaxy power spectrum through spectroscopic measurements rely on a number of assumptions (see list in \S~\ref{ssec:3D}) that will hold no longer if we analyse such new catalogues as a whole. A more correct approach is that of 2D tomography, which, however, either does not fully exploit the exquisite radial resolution of spectroscopic redshifts, or becomes computationally untreatable.

Thus, the method we have introduced here for the first time draws inspiration from the standard 3D $P_{\rm g}({\bm k},z)$ approach, but implements it within the consistent, gauge-independent formalism of 2D tomographic $C^{ij}_\ell$. For this reason, we refer to it as `hybrid'. Note that our method intentionally neglects lensing, in order to resemble more closely the standard method in observations. Our main results can be summarised as follows:
\begin{enumerate}
\item[\textit{(i)}] As presented in \S~\ref{ssec:constraints}, forecasts show that hybrid constraints on cosmological parameters are comparable to those from a $P_{\rm g}({\bm k},z)$ analysis, and more than twice as tight as those obtained via standard tomography, despite the larger number of nuisance parameters included. This trend is even more pronounced when limiting the analysis to strictly linear scales. We have also shown this by means of the increment of information gain, which roughly corresponds to that achieved by BAO measurements over WMAP9 data.
\item[\textit{(ii)}] Our hybrid approach is more robust than 2D tomography with respect to the neglect of corrections to galaxy number count fluctuations beyond density and RSD. In \S~\ref{ssec:biases}, we have shown that if we do not include lensing magnification (the most important remaining contribution), best fits on cosmological parameters stay within a few tens of percent of a standard deviation from the true values, whereas with a standard tomographic approach they get biased by more than $1\sigma$.
\end{enumerate}

Summarising, our method is able to match the $P_{\rm g}({\bm k},z)$ constraints on \lcdm\ cosmological parameters, while easily incorporating cosmic evolution, wide-angle effects and lensing effects. It does not require Alcock-Paczynski corrections, since the angular power spectrum is an observable and does not need one to assume a fiducial cosmology. It improves significantly on broad-bin 2D tomography. And it is not biased by neglecting lensing effects, i.e.\ even when implementing only the two main contributions to galaxy number count fluctuations, viz.\ density perturbations and RSD. This last fact leads to a further, major advantage compared to broad-bin tomography: it allows for faster code implementations, since the inclusion of integrated lensing terms significantly slows computation of angular power spectra. {In a forthcoming work we plan to estimate how the thickness of the redshift slicing affects the robustness of our method, namely what is the border between when it is safe to neglect lensing and when one has to include it.}

\section*{Acknowledgements}
We warmly thank Ruth Durrer for valuable comments, and Marco Raveri and Matteo Martinelli for insightful discussions about information gain. SC is supported by the Italian Ministry of Education, University and Research (MIUR) through Rita Levi Montalcini project `\textsc{prometheus} -- Probing and Relating Observables with Multi-wavelength Experiments To Help Enlightening the Universe's Structure', and by the `Departments of Excellence 2018-2022' Grant awarded by MIUR (L.\ 232/2016). SC also acknowledges support from ERC Starting Grant No. 280127 during the development of this project. We acknowledge the support of the Centre for High Performance Computing, South Africa, under the project ASTR0945. JF, RM and MGS are supported by the South African Square Kilometre Array Project and National Research Foundation (Grant Nos. 75415 and 84156). RM is also supported by the UK Science \&\ Technology Facilities Council (Grant No. ST/K0090X/1).

\bibliographystyle{mnras}
\bibliography{Bibliography,main_bib}

\appendix
\section{Fisher matrix formalism and tools}\label{app:Fisher}
We start from the assumption that  a given observable, for which we have a prediction based on some theoretical model, corresponds to a function of a  set of parameters $\boldsymbol\vartheta$. 
In the Fisher information matrix formalism, the observed outcome is the mean value of the observable assumed as the null hypothesis. Thanks to this, we can estimate errors on the parameters, given errors in observable quantities.

In the frequentist approach, the Fisher matrix $\mathbfss F$ is defined as the expectation value of the Hessian of the log-likelihood function $\mathcal L=-\ln L$, i.e.
\begin{equation}
\mathbfss F\left(\vartheta_\alpha,\vartheta_\beta\right)=\left\langle\frac{\partial^2\mathcal L}{\partial\vartheta_\alpha\partial\vartheta_\beta}\right\rangle,
\end{equation}
while in the Bayesian approach the data is no longer represented by random variables, and no averaging takes place. Hence the Fisher matrix is simply evaluated at the parameter maximum-likelihood best fit, namely
\begin{equation}
\mathbfss F\left(\vartheta_\alpha,\vartheta_\beta\right)=\left.\frac{\partial^2\mathcal L}{\partial\vartheta_\alpha\partial\vartheta_\beta}\right|_{\boldsymbol\vartheta=\overline{\boldsymbol\vartheta}}.
\end{equation}
The two definitions coincide if the data is Gaussian and the parameters enter the mean and the variance in a linear way, or in the case of forecasting \citep{Sellentin:2014zta}.

The Cramer-Rao inequality states that a model parameter $\vartheta_\alpha$ cannot have a variance smaller than $\delta(\vartheta_\alpha)=1/\sqrt{\mathbfss F(\vartheta_\alpha,\vartheta_\alpha)}$, when all other parameters are fixed (these are called `conditional errors'), or be measured to a precision better than $\sigma(\vartheta_\alpha)=[\mathbfss F^{-1}(\vartheta_\alpha,\vartheta_\alpha)]^{1/2}$, when all other parameters are marginalised over (`marginal errors'). Here, $\mathbfss F^{-1}$ denotes the inverse of the Fisher matrix.

\subsection{Stability tests on Fisher matrices}\label{ssec:marginalisation}
To assess the validity of our results, we proceed as follows. We note that, particularly in the presence of nuisance parameters, the dimensionality of the Fisher matrices is high and their eigenvalues span a wide dynamical range. This may lead to numerical instabilities in the inversion of the Fisher matrix, and thus to spurious constraints on the cosmological parameters of interest. Therefore, to reduce the dynamic range, we first change the Fisher matrix basis via
\begin{equation}
\mathbfss F\left(\vartheta_{\alpha^\prime},\vartheta_{\beta^\prime}\right)=\mathbfss J^T_{\alpha^\prime\alpha}\mathbfss F\left(\vartheta_\alpha,\vartheta_\beta\right)\mathbfss J_{\beta^\prime\beta},\label{eq:F_jacobian}
\end{equation}
where $\mathbfss J_{\alpha^\prime\alpha}$ is the Jacobian of the transformation from $\vartheta_\alpha$ to $\vartheta{_\alpha^\prime}$. As a new basis we choose $\vartheta{_\alpha^\prime}=\big\{\ln\ob,\ln\odm,\ln\ns,\As,\ln h\big\}$, thus reducing the spread between the maximum and minimum eigenvalues by a factor of $4.3\times10^4$.

We marginalise over the bias nuisance parameters, 20  for the standard binning approach and $10\times14$ for our new hybrid approach. The marginalisation involves a first inversion of the Fisher matrix, whose rows and columns referring to the bias nuisance parameters are then dropped. At this point, the inverse of the Fisher matrix (i.e.\ the parameter covariance matrix) only contains entries relative to the cosmological parameters. Then we can invert it  and obtain the marginalised Fisher matrix in the new basis. Lastly, we perform the inverse change of variable of Eq.~\eqref{eq:F_jacobian}, thus obtaining the final, marginalised Fisher matrix for the initial cosmological parameter set.
\begin{table*}
\centering
\caption{Maximum discrepancy between the inverse of the Fisher matrix, $\mathbfss F^{-1}$, and its Moore-Penrose pseudo-inverse, $\mathbfss F_{\rm MP}^{-1}$, or its inverse via eigen-decomposition, $\mathbfss F_{\rm ED}^{-1}$.}
\begin{tabular}{lrr}
& $\max\left\{\left|\mathbfss F_{\rm MP}^{-1}/\mathbfss F^{-1}-1\right|\right\}$ & $\max\left\{\left|\mathbfss F_{\rm ED}^{-1}/\mathbfss F^{-1}-1\right|\right\}$\\
\hline
2D without $A_\kappa$ & $1.3\times10^{-10}$ & $4.6\times10^{-7}$ \\
2D with $A_\kappa$ & $8.9\times10^{-10}$ & $1.4\times10^{-6}$ \\
hybrid without $A_\kappa$ & $5.0\times10^{-10}$ & $1.5\times10^{-4}$ \\
hybrid with $A_\kappa$ & $4.3\times10^{-11}$ & $5.7\times10^{-6}$ \\
\end{tabular}\label{tab:F_inv}
\end{table*}

Despite the reduction of the dynamic range, matrix inversions may still propagate numerical instabilities into the resulting Fisher matrix. Therefore, we check the stability of our inverse matrices $\mathbfss F^{-1}$ (computed via one-step row reduction) by comparing them to the Moore-Penrose pseudo-inverse and to the inverse obtained via eigen-decomposition. This last approach helps in removing degeneracies from the Fisher matrix. Indeed, when marginalising over the set of nuisance parameters, if one or more eigenvalues (nearly) vanish, then this degeneracy does not propagate into the cosmological parameters of interest \citep{Albrecht:2006um,Camera:2012ez}. For a well-defined square matrix, the three matrix inversions must coincide. If we denote the Moore-Penrose pseudo-inverse of $\mathbfss F$ by $\mathbfss F_{\rm MP}^{-1}$ and the eigen-decomposed inverse by $\mathbfss F_{\rm ED}^{-1}$, we can quantify the reliability of the marginal Fisher inversion by computing $\big|\mathbfss F_X^{-1}/\mathbfss F^{-1}-1\big|$, with $X=\mathrm{MP},\mathrm{ED}$. Table~\ref{tab:F_inv} shows the maximum values of $\big|\mathbfss F_X^{-1}/\mathbfss F^{-1}-1\big|$ for the  marginal Fisher matrices employed in this work. It is easy to see that the discrepancy between the various methods of matrix inversion is at the very most $\mathcal O(0.01\%)$. We conclude that our marginal Fisher matrices are robust under inversion, and so are our cosmological parameter constraints.

\subsection{Estimating the bias on cosmological parameters}\label{ssec:bias}
Given a set of cosmological, nuisance and systematic-effect parameters, we can estimate the amount of biasing we will incur if we neglect some systematic effect within the Fisher information matrix framework. If
\begin{equation}
\mathbfcal F=\left[
\begin{array}{c|c}
\mathbfss F\left(\vartheta_\alpha,\vartheta_\beta\right) & \mathbfss F\left(\vartheta_\alpha,\varphi_b\right)\\
\hline
\mathbfss F\left(\varphi_a,\vartheta_\beta\right) & \mathbfss F\left(\varphi_a,\varphi_b\right)
\end{array}
\right]
\end{equation}
is the full Fisher matrix, marginalised over nuisance parameters, that includes both cosmological parameters, $\boldsymbol\vartheta$, and systematic-effect  parameters, $\boldsymbol\varphi$, we can compute the bias as (\citealp{Heavens:2007ka}; see also \citealp{Camera:2016owj}, Appendix A)
\be
b(\vartheta_\alpha)=\mathbfcal F^{-1}\left(\vartheta_\alpha,\vartheta_\beta\right)\mathbfcal F\left(\vartheta_\beta,\varphi_a\right)\delta\varphi_a,\label{eq:bias}
\ee
where $\mathbfcal F^{-1}(\vartheta_\alpha,\vartheta_\beta)\ne[\mathbfcal F(\vartheta_\alpha,\vartheta_\beta)]^{-1}$ means that one has first to invert $\mathbfcal F$ and then consider only the rows and columns of the \textit{full} Fisher matrix relative to the cosmological parameters $\boldsymbol\vartheta$, whilst $\mathbfcal F(\vartheta_\beta,\varphi_a)\equiv\mathbfss F(\vartheta_\beta,\varphi_a)$. Here, $\delta\varphi_a$ are the shifts from the true values of $\varphi_a$ to the values incorrectly assumed in the analysis.

\subsection{Information gain}\label{sapp:IG}
As a figure of merit to compare the new method we outlined in \S~\ref{ssec:2.5D}, we introduce here the concept of the Kullback-Leibler divergence, also known as `relative entropy' or `information gain' \citep{Kullback:1951}. This is a tool for multivariate posterior distributions to compare the constraining power from different data sets or survey implementations. If ${\bm x}$ is a continuous, $d$-dimensional random variable with probability density functions $p_1({\bm x})$ and $p_2({\bm x})$, the information gain is
\begin{equation}
D(p_2||p_1)=\frac{1}{2\ln2}\int\!\!\de^dX\,p_2({\bm x})\ln\frac{p_2({\bm x})}{p_1({\bm x})}\,\mathrm{bits}\,.
\end{equation}
It represents the information gain (in bits) obtained by updating the distribution describing ${\bm x}$ from $p_1$ to $p_2$. Although it is not symmetric in $p_2$ and $p_1$, $D(p_2||p_1)$ is often interpreted as a distance between the two distributions. Indeed, it has remarkable properties, such as being non-negative, $D(p_2||p_1)\ge0$, and zero if and only if $p_2=p_1$.

In the case of multivariate normal (Gaussian) posterior distributions $\mathcal N(\boldsymbol\vartheta)$ for the cosmological parameters, the information gain from a prior knowledge $\mathcal N_1$ to the posterior obtained from a new experiment $\mathcal N_2$ is
\begin{equation}
D(\mathcal N_2||\mathcal N_1)=\frac{1}{2\ln2}\left\{\ln\frac{\det\Sigma_1}{\det\Sigma_2}+{\rm tr}\left[\Sigma_1^{-1}\left(\Sigma_2-\Sigma_1\right)\right]\right\}\,\mathrm{bits}\,,\label{eq:IG_Gaussian}
\end{equation}
where $\Sigma_i$ denotes the covariance matrix of the multivariate Gaussian $\mathcal N_i(\boldsymbol\vartheta)$. Note that we consider the prior and posterior distributions as having the same mean. Even though this is not usually the case with real data, it is a standard ansatz in forecasts. $\mathcal N_2$ can either be interpreted as the posterior of the data from a completely new experiment/analysis with respect to a previous experiment/analysis $\mathcal N_1$, or an updated posterior over a prior represented by $\mathcal N_1$.

\bsp	
\label{lastpage}
\end{document}